\documentclass[prb,rsi,amsmath,amssymb,reprint, onecolumn]{revtex4-1}

\usepackage{graphicx}
\usepackage{dcolumn}
\usepackage{bm}
\usepackage{natbib}
\setcitestyle{square}

\usepackage{amsmath}
\usepackage{braket}

\usepackage{mathrsfs}
\usepackage{amsfonts}
\usepackage{color}
\usepackage[normalem]{ulem} 
\usepackage{longtable}
\usepackage{cellspace}
\usepackage{hyperref}
\usepackage{siunitx}

\hypersetup{
    colorlinks=true,
    linkcolor=blue,
    filecolor=magenta,      
    urlcolor=black,
    citecolor=red,
}

\begin{document}

%\preprint{APS/123-QED}

\title{Revisiting the Thomas-Fermi Potential for Three-Dimensional Condensed Matter Systems}
\author{Gionni Marchetti}
\email{gionnimarchetti@ub.edu}
\email{gionnimarchetti@gmail.com}

\affiliation{%
Departament de F\'{i}sica de la Mat\`{e}ria Condensada, Facultat de Física,  Universitat de Barcelona, Carrer Mart\'{i} i Franqu\`{e}s 1, 08028, Barcelona, Spain\\
Institut de Nanoci{\`e}ncia i Nanotecnologia, Universitat de Barcelona, Av. Joan XXIII S/N, 08028, Barcelona, Spain}

\date{\today}

\begin{abstract}
We proposed a formally exact, probabilistic method to assess the validity of the Thomas-Fermi potential for three-dimensional condensed matter systems where electron dynamics is constrained to the Fermi surface. Our method, which relies on accurate solutions of the radial Schr\"{o}dinger equation, yields the probability density function for momentum transfer. This allows for the computation of its expectation values, which can be compared with unity to confirm the validity of the Thomas-Fermi approximation. We applied this method to three {\it n}-type  direct-gap III-V model semiconductors (GaAs, InAs, InSb) and found that the Thomas-Fermi approximation is certainly valid at high electron densities. In these cases, the probability density function exhibits the same profile, irrespective of the material under scrutiny. Furthermore, we show that this approximation can lead to serious errors in the computation of observables when applied to GaAs at zero temperature for most electron densities under scrutiny.
 \end{abstract}

\maketitle

\section{Introduction} \label{intro}

The random phase approximation (RPA) is arguably one of the most important mean-field theories in condensed matter theory and materials science~\cite{ bohm1953,  pines2016, ren2012}. This approximation is particularly effective in   modelling the  screening  in  free electron-like materials~\cite{penn1987, enfietzoglou2013} and is also a crucial ingredient for the GW approximation~\cite{hedin1965,  golze2019, Marzari2021}. However, in some of its applications  
the full static interaction at RPA level, in the following denoted by  $V_{\rm ei}^{\rm RPA}$, is neglected and replaced by a very crude approximation: the Thomas-Fermi (TF) potential $V_{\rm ei}^{\rm TF}$~\cite{ashcroft1976}. To illustrate this fact, we recall that the TF potential is widely used for computing some important physical quantities such as the electron lifetime (also referred to as the single-particle relaxation time) in  materials~\cite{dasSarma1985, caruso2016}, the electron and hole mobilities in semiconductors~\cite{hall1962, chattopadhyay1981, meyer1981, meyer1981a}, the screened second order M{\o}ller-Plesset  amplitude~\cite{shepherd2013} and the effects of charged impurity scattering on the transport properties  of graphene at zero temperature~\cite{hwang2007}. 

\begin{figure}

\resizebox{0.50\textwidth}{!}{%
  \includegraphics{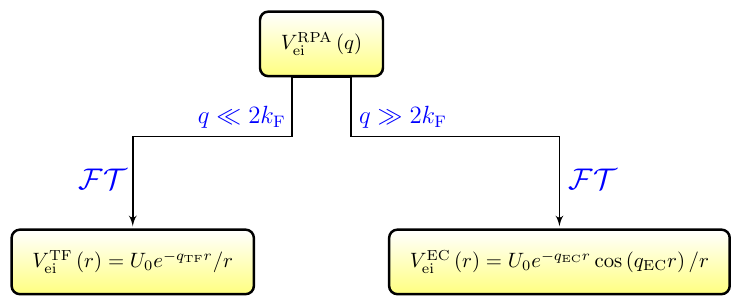}
}

\caption{The analytical potentials $V_{\rm ei}^{\rm TF}\left(r\right)$ and $V_{\rm ei}^{\rm EC}\left(r\right)$ in coordinate space $r$, obtained by expanding $V_{\rm ei}^{\rm RPA}\left(q\right)$ in the momentum space $q$ for the small ($q \ll 2 k_{\rm F}$) and large ($q \gg  2 k_{\rm F}$) momentum transfer $q$ with respect to the Fermi wave number  $k_{\rm F} $, and thereafter performing a standard Fourier transform ($\mathcal{FT}$)~ \cite{ashcroft1976}. The symbol $U_0$ denotes the  potential's strength. We refer the reader to Section~\ref{rpa} for the definitions of the constants $q_{\rm TF}$ and $q_{\rm EC}$.}
\label{fig:figure1}      
\end{figure}

Generally, researchers do not bother providing a physics-based explanation for using such a crude potential. Therefore, one might expect that this choice is primarily dictated by the  potential's simple functional form. However, if this were the case, it would be quite surprising because at the RPA level, there is another analytical potential available: the exponential cosine  (EC) screened Coulomb potential $V_{\rm ei}^{\rm EC}$, that also takes a simple functional form~\cite{takimoto1959, bonch-bruevich1962, lam1972} (see cartoon of Fig.~\ref{fig:figure1} for their comparison). It is worth noting here that the EC potential is widely employed for modelling the screened Coulomb interaction in an ideal quantum plasma~\cite{shukla2008, shukla2012, nayek2012, qi2016, janev2016, munjal2017, chen2023} despite the fact that  this potential is another crude approximation of RPA interaction. In fact,  both the TF and EC potentials lack a significant portion of the electronic structure information contained in the Lindhard function, as discussed in detail in  Section~\ref{rpa}.

That being said, there exists a genuine need for a rigorous method to assess the validity of the chosen approximate RPA potential, thereby avoiding erroneous computations of observables of interest. In this paper, we shall present a formally exact probabilistic method that can enable researchers to determine the consistency of the Thomas-Fermi potential in a rigorous manner. Such a method is based upon  two important facts: the fermionic dynamics is constrained to the Fermi surface (FS)~\cite{polchinski1999effective} and the TF arises from the  long-wave limit expansion of the Lindhard function, that is, for  $q \ll 2 k_{\rm F}$ where $q$, $k_{\rm F}$  denote the wave number and  Fermi wave number, respectively. Now, the wave vector (or scattering vector) $\mathbf{q}$ can be physically understood as the momentum transferred due to the exchange of a photon in electron-test-charge scattering, which occurs during the fermionic dynamics on FS. As a result, one can derive the probability density function (PDF)  $f_x \left(x\right)$ ($x\equiv q/2k_{\rm F}$) of momentum transfer (see Eq.~\ref{eq:pdf5}) according to the scattering theory. By means of  $f_x$ is then possible to compute the expectation value $\Bar{x}$ of the random variable $x$, thereby confirming the validity of TF potential  applied to a given system  whenever $\Bar{x} \ll 1$. As explained in Section~\ref{pdf}  the PDF is built on  quantum scattering phase shifts $\delta_l$ which depend on the  details of screening~\cite{bethe1957} and as consequence one needs to accurately solve the radial  Schr\"{o}dinger  equation for the TF potential. To this end, we shall employ the variable phase method (VPM)~\cite{calogero1963, calogero1967, babikov1967} which yields very accurate phase shifts at inexpensive computational cost.

In the following, we shall apply our principled method  to three {\it n}-type direct gap III-V model semiconductors with a zinc blende structure: Gallium Arsenide (GaAs), Indium Arsenide (InAs) and Indium Antimonide (InSb).  We have chosen them for their different electron effective mass and the dielectric background constant, as shown in Table~\ref{table:table1}. Furthermore, it is possible to tune their electron density, denoted  as $n$, and, hence, the screening parameter  characterizing the Thomas-Fermi potential, which depends on electron density and  other material parameters.

 \begin{table}
\renewcommand{\arraystretch}{2} 
\caption{The effective mass $m^{\ast}$ and the relative permittivity  $\kappa$ of III-V semiconductors GaAs, InAs, and InSb with zinc blend structure used for the computations. The data are taken from Refs.~\cite{madelung1991, vurgaftman2001}. }
\label{table:table1}
  \begin{ruledtabular}
    \begin{tabular}{c | c   c   c  }
    Parameters & GaAs & InAs  & InSb  \\ [1ex]
    \hline
    $m^{\ast}/m $     &  $0.067$       & $0.026 $        & $0.014 $ \\

    $\kappa$    &     $12.9 $    & $15.15$         & $16.8 $       \\ [1ex]
    \end{tabular}
\end{ruledtabular}
\end{table}
% ----------------------------------------------
%

Our findings clearly demonstrates that,  the Thomas-Fermi potential is certainly applicable in the high-density limit as one would expected being derived from the random phase approximation. In this context, we have observed that the probability density function follows a common pattern regardless of the material under investigation, yielding an approximate expectation value of $\Bar{x} \approx 0.3$. This is a direct consequence of the smallness of the scattering phase shifts in the metallic regime. Furthermore, our analysis reveals that the Thomas-Fermi approximation holds even at low and intermediate electron densities for InAs and InSb. However, applying it to GaAs at zero temperature would likely result in highly questionable outcomes.

\section{Review of the Analytical Potentials from the full RPA Interaction} \label{rpa}

Within the self-consistent field approximation, the screened Coulomb interaction potential $V_{\rm ei}^{\rm RPA}$ between a positive test-charge with charge $Z$ (in units of the elementary charge $e$) and an electron in the static limit ($\omega \to 0$)  reads~\cite{giuliani2005, schliemann2011} 

\begin{equation}\label{eq:rpa1}
V_{\rm ei}^{\rm RPA} \left(r\right) = \frac{1}{\left(2 \pi\right)^{3}}\int d^{3}q 
\frac{V_{ei}\left(\mathbf{q} \right)}{\varepsilon^{\rm RPA}\left(\mathbf{q}, \omega =0 \right)}  e^{i \mathbf{q}\cdot \mathbf{r}} \, ,
\end{equation}
where  $V_{\rm ei}= U_0 r^{-1}$ with $U_0= - Ze^2/\kappa $ (Gaussian units) is the bare Coulomb interaction in coordinate space, and  $\kappa$ denotes the background static dielectric constant. Eq.~\ref{eq:rpa1} assumes that the presence of a static test-charge causes  small perturbations to the homogeneous  electron gas, thereby making the linear response theory applicable. In  Eq.~\ref{eq:rpa1} the static dielectric function  $\varepsilon^{\rm RPA}\left(\mathbf{q}\right) \equiv \varepsilon^{\rm RPA}\left(\mathbf{q}, \omega=0 \right) $ reads

\begin{equation}\label{eq:rpa2}
\varepsilon^{\rm RPA}\left(\mathbf{q}\right) = 1 - v_q \left(\mathbf{q}\right)\Pi^{0} \left(\mathbf{q}\right) \, ,
\end{equation}
where $v_q \left(\mathbf{q}\right)= 4 \pi e^2/\kappa q^2 $ ($q = |\mathbf{q}|$) and $\Pi^{0}$ denotes the free polarizability  of a three-dimensional (3D) electron gas \cite{lindhard1954, ashcroft1976}. Recalling that at zero temperature, the density of states per unit volume at the Fermi energy  of a 3D electron gas in a paramagnetic state is  $N(0)\equiv m^{\ast}k_{\rm F}/\pi^{2}\hbar^{2}$ where $m^{\ast}$ denotes the electron's effective mass ($\hbar= h/2\pi $ where $h$ is the Planck constant), one can write $\Pi^{0} \left(x\right) = N(0) G\left(x\right)$ where the  Lindhard function $G \left(x\right) $ is usually defined  in terms of the dimensionless variable $x \equiv q/2k_{\rm F}$ as\cite{ashcroft1976}

\begin{equation}\label{eq:rpa3}
G \left(x\right) = \frac{1}{2 } +  \frac{1 - x^2}{ 4 x } \ln{\left\lvert \frac { 1 + x}{1 - x}\right\rvert} \, .
\end{equation}

Note that the free polarizability corresponds to the bare bubble diagram~\cite{giuliani2005}. Furthermore, it would be expected that the RPA is asymptotically exact in the high-density limit \cite{gellmann1957}. This means that the density parameter (or Wigner-Seitz radius~\cite{giuliani2005}) $r_s$ ($r_s= \left(3/4\pi  n \right)^{1/3}$ when $n$ is expressed in atomic units) must be smaller than unity, i.e. $r_s \ll 1$. However, the above inequality loosely holds for many condensed matter systems to which this approximation is applied. In fact, the RPA proves to be a successful approximation in two-dimensional semiconductors and metals despite the fact that $r_s \approx 2-10$~\cite{adam2007}.

\begin{figure}

\resizebox{0.50\textwidth}{!}{%
  \includegraphics{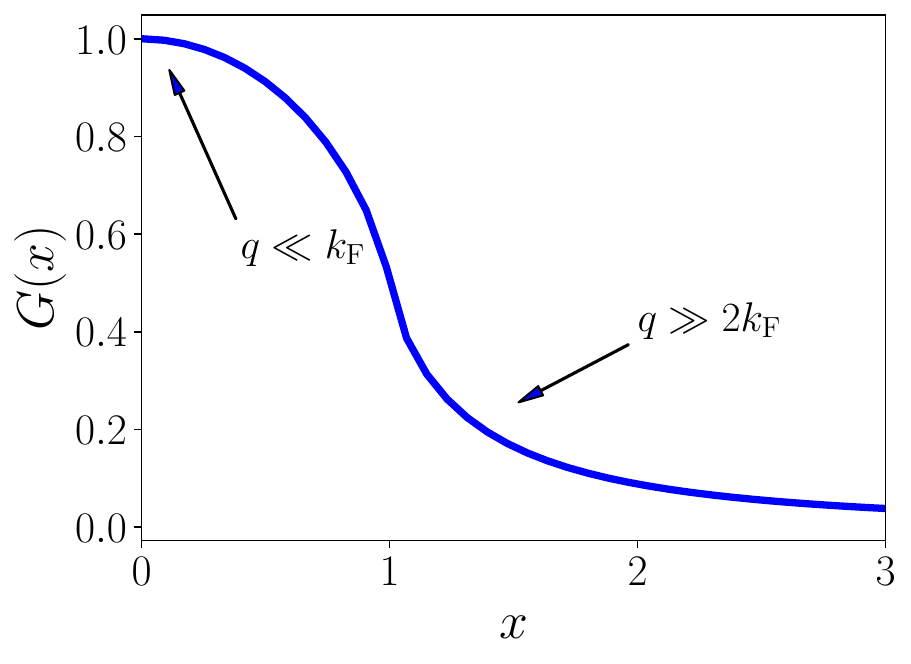}
}

\caption{The Lindhard function $G \left(x\right) $ as function of $x= q/2k_{\rm F}$. The arrows denote the regions  where 
the expansions, see Eqs. \ref{eq:rpa4}, \ref{eq:rpa5},  are expected to be valid, that is, in regions far from the logarithmic singularity due to the existence of a sharp Fermi surface  at $x=1$.  }
\label{fig:figure2}      
\end{figure}

The Thomas-Fermi and exponential cosine screened Coulomb potentials arise
from expansions of $G\left(q\right)$ in small and  large wave number $q$ regions, respectively,~\cite{takimoto1959, langer1960, hall1962} as shown in Fig.~\ref{fig:figure2}.  In this regard, one finds that

\begin{equation}\label{eq:rpa4}
G \left(x\right) \approx 1 - \frac{x^2}{3} - \frac{x^4}{15} + \cdots \qquad\text{for $x \ll 1 $} \, ,
\end{equation}
and 
\begin{equation}\label{eq:rpa5}
G \left(x\right) \approx  \frac{x^{-2}}{3} + \frac{x^{-4}}{15 } + \cdots 
\qquad\text{for $x \gg 1 $} \, .
\end{equation}

Next, inserting Eqs.~\ref{eq:rpa4}, \ref{eq:rpa5}   into Eq.~ \ref{eq:rpa1}, and thereafter performing the Fourier transform according to the standard definition given in Ref.~\cite{ashcroft1976},  one gets  $V_{\rm ei}^{\rm TF} \left(r\right)=  -\left(Ze^{2}/\kappa r\right)e^{- q_{\rm TF} r} $ and  $V_{\rm ei}^{\rm EC} \left(r\right)= -\left(Ze^{2}/\kappa r\right) e^{- q_{\rm EC} r} \cos(q_{\rm EC} r) $, respectively. The screening parameters are $q_{\rm TF} =  \left(6 \pi e^{2}n/E_{\rm F}\right)^{1/2}$ and 
$q_{\rm EC} =  \left(m^{\ast}\omega_{\mathrm{\rm pe}}/\hbar\right)^{1/2}$, ~\cite{ashcroft1976,mcIrvine1960, shukla2008}, respectively.  In the above parameters,  $E_{\mathrm{F}}= \hbar^{2} k_{\mathrm{F}}^{2}/2 m^{\ast} $ and  $\omega_{\mathrm{\rm pe}}$ denote the Fermi energy and the electron plasma frequency~\cite{shukla2008, giuliani2005}, respectively. 

In general, as the RPA is equivalent to  a time-dependent Hartree approximation~\cite{giuliani2005}, these two approximate potentials cannot account for the short-range exchange and correlation effects. According to Ref.~\cite{bonch-bruevich1962}, the EC potential is expected to account for the quantum correlation effects at small distances ($r  \ll 1/k_{\rm F}$ ) from the test-charge. These small distances define  the so-called \emph{ultra-quantum} region~\cite{bonch-bruevich1962}. On the other hand, the 
the TF potential can be also obtained on a semiclassical basis followed by  a subsequent appropriate linearization procedure~\cite{ashcroft1976}. This alternative derivation suggests that the TF potential might miss significant  quantum effects due to its semiclassical origin. Furthermore, it is worth noting here that in the literature the Thomas-Fermi potential is often derived assuming a stronger approximation, that is, $q \ll k_{\rm F}$~\cite{ashcroft1976, giuliani2005}, than the one defining  the expansion in Eq.~\ref{eq:rpa4}.

Finally, we conclude by noting that both of these potentials are rather crude approximations of the full interaction potential. In fact, neither the Thomas-Fermi potential nor the exponential cosine screened potential can accurately capture the density oscillations~\cite{hohenberg1964}, e.g. the Friedel oscillations~\cite{friedel1952}, because their corresponding expansions are performed in regions far away from the Lindhard function's singularity at $q=2 k_{\rm F}$.

\section{Derivation of the Momentum Transfer Probability Density Function} \label{pdf}

In the following, we shall derive the momentum transfer probability density function  $f_{x}$ and apply it to the TF potential. Our aim is to determine the average  wave number transfer $  \Bar{x}$ by means of $f_{x}$ for the material's parameters under scrutiny.  Subsequently, we would expect the consistency of TF potential  to be reached whenever    $ \Bar{x} \ll 1 $.

The momentum transfer is defined as $\mathbf{q} = \mathbf{k} -\mathbf{k'} $  where $\mathbf{k}$ and $\mathbf{k}^{\prime}$  denote the electron's plane wave vectors  before and after a collision with a test-charge, respectively. Note that throughout we shall assume  $Z=1$.
Because the fermionic dynamics is constrained on the Fermi surface, one can assume that $k=k'=k_{\mathrm{F}}$ and  $q = 2 k_{\mathrm{F}} \sin\left(\theta/2\right)$ where $\theta$ ($\theta \in [0, \pi]$) is the respective scattering angle. In order to construct the PDF we need the  differential scattering cross-section  $d \sigma^{\rm  ei} \left(\theta, \phi \right)/ d \Omega $~\cite{stefanucci2013} that yields the  probability that an electron is scattered off into the infinitesimal solid angle $d \Omega =  \sin\theta d\theta d\phi $.
 However, due to the cylindrical symmetry of the problem at hand, the differential cross-section  will not depend upon the  azimuthal angle $\phi \in [0, 2\pi]$. Note that this fact justifies the so-called central-field approximation of the atomistic systems~\cite{schiff1968}. Accordingly, the probability density function  $f_{\rm \theta}$ will be  a function  of the (random) variable $\theta$ only. As a consequence,  $f_{\rm \theta}$ is a one-dimensional function defined on the domain $ [0, \pi]$ that reads~\cite{lundstrom2000} 

\begin{equation}\label{eq:pdf1}
f_{\rm \theta} \left(\theta \right)    = \frac{ \left(d\sigma^{\rm  ei} \left(\theta \right) /d \Omega \right) \sin\theta }{\int_0^\pi d \, \theta \left(d\sigma^{\rm  ei} \left(\theta \right)/d \Omega \right) \sin\theta  }  \, .
\end{equation}

%$d\, \theta$

In the present work, the function $f_{\theta}$ yields the probability that an electron with Fermi energy will be scattered into an angle between $\theta$ and $\theta + d \theta$. Note that the integral at the denominator  of Eq.~\ref{eq:pdf1}, is the normalization factor, that in turn is proportional the electron-test-charge total cross-section.

Next, we shall expand $d \sigma^{\rm  ei} / d \Omega $  in polynomials of Legendre $P_l \left(\cos\theta  \right)$. Then, one finds~\cite{schiff1968}

\begin{equation}\label{eq:legendre}
 \frac{d\sigma^{\rm  ei}   }{d \Omega } = \frac{1}{k_{\rm F}^{2}}
\sum_{l=0}^{\infty}\sum_{l'=0}^{\infty}\left(2l+1\right)\left(2l'+1\right)
\sin\delta_l\sin\delta_l' P_{l}\left(\cos\theta\right) P_{l'}\left(\cos\theta\right)\, ,
\end{equation}
where $\delta_l$ denotes the scattering phase shift corresponding to the angular momentum number $l$. In this work we shall  compute the quantum phase shift by accurately solving the radial   Schr\"{o}dinger equation for  Thomas-Fermi potential $ V^{\rm TF}_{\rm ei}$ through   the variable phase method~\cite{calogero1963, babikov1967, calogero1967, marchetti2019}. According to this method the phase functions $\delta_l\left(r\right)$  are  the solutions of the following first order nonlinear differential equation (a generalized Riccati equation)
\begin{equation}\label{eq:phase}
  \delta'_l \left(r \right) = - \frac{2  m^{\ast}   V^{\rm TF}_{\rm ei}\left(r \right) }{ k_{\rm F} \hbar^{2}}  \left[\cos\delta_l\left(r \right)	\hat{j_l} \left(kr \right)  - \sin\delta_l\left(r \right)\hat{n}_l \left(kr \right)  \right]^{2}\, , 
\end{equation}
with the initial condition at the origin: $\delta_l\left(0\right)=0$. Note that in Eq.~\ref{eq:phase} $\hat{j_l}$, $\hat{n}_l$ are the Riccati-Bessel functions~\cite{calogero1963}.

Here, we shall limit ourselves to the first three partial wave contributions, i.e., those corresponding to $l=0, 1, 2$, to the probability density function. This choice prevents the PDF formulas from becoming overly cumbersome. However, additional refinements are possible. Adding new contributions is certainly not computationally costly, but it generally requires some care. Indeed, it is essential to ensure that the formulas yield a non-negative function, as expected for a PDF, across its entire domain.

That being said, combining equations \ref{eq:pdf1} and \ref{eq:legendre},  we obtain the following expression for  $f_{\theta}\left(\theta \right)$

\begin{widetext}

\begin{equation}\label{eq:pdf2}
f_{\theta}\left(\theta \right) \approx \frac{1 }{N_{\theta}}\Bigg[a_1 + a_2\cos\theta  + a_3\cos^{2}\theta  + a_4 \left(3 \cos^{2}\theta - 1 \right)  + a_5 \cos\theta \left(3 \cos^{2}\theta - 1 \right)  + a_6  \left(3 \cos^{2}\theta - 1 \right)^{2}  \Bigg]\sin\theta  \, ,
\end{equation}

\end{widetext}
where $a_1 =\sin^{2} \delta_0 $,  $a_2 = 6 \sin \delta_0 \sin\delta_1\cos\left(\delta_0 - \delta_1\right) $,  $a_3 = 9\sin^{2} \delta_1$, $a_4 = 5 \sin \delta_0 \sin\delta_2\cos\left(\delta_0 - \delta_2\right) $, 
$a_4 = \sin^{2} \delta_1$,  $ a_5 = 15 \sin \delta_1 \sin\delta_2\cos\left(\delta_1 - \delta_2\right) $ and $a_6 = \left(25/4\right) \sin^{2} \delta_2 $.  In Eq.~\ref{eq:pdf2}  $N_{\theta}$ denotes the appropriate normalisation factor.

Next, the momentum transfer probability density function $f_x$ can be derived directly from $f_{\theta}$ through  the change-of-variable formula~\cite{Deisenroth2020}. In fact, defining the function $g$
such that $x \equiv g\left(\theta\right) = \sin\left(\theta/2\right) $ and noting that it is a monotonically increasing function mapping the interval $[0,\pi]$ to $[0,1]$, one can apply the following change-of-variable formula

\begin{equation}\label{eq:pdf3}
f_{x}\left(x \right) = f_{\theta} \left(g^{-1}\left(x \right) \right)\left| \frac{d  }{d x} g^{-1}\left(x \right)\right|
\, ,
\end{equation}
where $g^{-1}$ denotes the inverse function of $g$. Using the fact, that
 
\begin{equation}\label{eq:pdf4}
 \frac{d  }{d x} g^{-1}\left( x\right)= \frac{2  }{\sqrt{1 - x^{2}}} 
\, ,
\end{equation}
one obtains the following PDF $f_{x}$ for the new variable $x$

 \begin{widetext}
\begin{equation}\label{eq:pdf5}
f_{x}\left(x \right) = \frac{4 x }{N_x}\Bigg[a_1 + a_2 \tilde{f}\left(x\right) + a_3 \tilde{f}^{2}\left(x\right) + a_4    \left( 3\tilde{f}^{2}\left(x\right) -1 \right)
+ a_5 \tilde{f}\left(x\right) \left( 3\tilde{f}^{2}\left(x\right) -1 \right) + a_6 \left( 3\tilde{f}^{2}\left(x\right) -1 \right) ^{2} \Bigg]\, ,
\end{equation}
\end{widetext}
where $N_x$ denotes the appropriate normalisation factor. In Eq.~\ref{eq:pdf5}, we introduced the function $\tilde{f}\left(x\right) = 1 - 2x^{2}$, that is equivalent to set $ \cos \theta  \equiv \tilde{f}\left(x\right) $.  
 
Finally, the  expected momentum transfer wave number 
 $ \Bar{x}$ can be numerically computed either by means of  the first moment of $f_x$, that is, 
   
\begin{equation}\label{eq:pdf6}
\Bar{x} = \int_0^1 dx \, x f_{x}\left(x \right)\, ,
\end{equation}
or with the help of the following analytical formula
\begin{equation}\label{eq:analytical1}
N_x = 2 a_1 + \frac{2}{3}a_3 + \frac{8}{5}a_6  \, ,
\end{equation}
and 
\begin{equation}\label{eq:analytical2}
 \Bar{x} = \frac{4}{N_x} \left(\frac{a_1}{3} - \frac{a_2}{15} + \frac{11a_3}{105} - \frac{2a_4}{105} - \frac{2a_5}{35} + \frac{292a_6}{1155} \right) \, .
\end{equation}

We found that both Eq.~\ref{eq:analytical1} and Eq.~\ref{eq:analytical2} are in excellent agreement.

\begin{table*}[tbp]
\renewcommand{\arraystretch}{2} 
\caption{Phase Shifts $\delta_0, \delta_1, \delta_2$ computed by numerically solving the radial  Schr\"{o}dinger  equation through Eq.~\ref{eq:phase} starting with initial condition at the origin $\delta_l\left(0\right) =0$  for each $l=0,1,2$ and assuming the wave vector $ k_{\rm F}$ as incoming wave number. The numbers in parentheses indicate the powers of $10$. Here $n, r_s, \Bar{x}$ corresponds to the electron density, the Wigner-Seitz radius and the expectation value of the probability density function, respectively. The expectation value $\Bar{x}$ is computed by means of Eq.~\ref{eq:analytical2}.}
\label{table:table2}
  \begin{ruledtabular}
    \begin{tabular}{c | c   c   c  c c c  }
   {\it n}-type semiconductor &  $n$  & $r_s$ &   $\delta_0$ &   $\delta_1$ &   $\delta_2$ &  $\Bar{x}$ \\ [1ex]
    \hline
    GaAs     & $10^{16}$ $\mathrm{cm}^{-3}$    & $2.83 $        & $1.36 $      & $1.90 \left(-1\right)$     & $3.69  \left(-2\right) $  & $0.59 $      
    \\
    
 -    &   $10^{18}$ $\mathrm{cm}^{-3}$   & $0.61 $        & $4.15  \left(-1\right)$      & $1.45 \left(-1\right) $     & $6.15 \left(-2\right) $  & $0.39$      \\
 
  -    &   $10^{21}$ $\mathrm{cm}^{-3}$   & $0.06 $        &  $7.33  \left(-2\right)$       &  $4.29  \left(-2\right)$    &  $2.90  \left(-2\right)$   & $0.31 $      \\
   \hline
   InAs     &  $10^{16}$ $\mathrm{cm}^{-3}$    & $0.93 $        & $5.68  \left(-1\right)$      & $1.66  \left(-1\right)$     & $5.96 \left(-2\right)$  & $0.43 $      \\
    -   &   $10^{18}$ $\mathrm{cm}^{-3}$   & $0.20 $        & $1.84\left(-1\right) $      & $8.85 \left(-2\right)$     & $4.50\left(-2\right) $  & $0.33 $      \\
   
 -   &   $10^{21}$ $\mathrm{cm}^{-3}$   & $0.02$        & $2.99 \left(-2\right) $      & $1.96 \left(-2\right)$     & $1.46 \left(-2\right) $  & $0.30 $      \\
\hline 
  InSb    &   $10^{16}$ $\mathrm{cm}^{-3}$    & $0.45 $        & $3.34  \left(-1\right) $      & $1.30  \left(-1\right)$     & $6.03  \left(-2\right)$  & $0.36 $  
  \\
  
    -   &  $10^{18}$ $\mathrm{cm}^{-3}$     & $0.09 $        & $1.06 \left(-1\right)$      & $5.82 \left(-2\right) $     & $3.70 \left(-2\right) $  & $0.31 $  
    \\
   
 -   &   $10^{21}$ $\mathrm{cm}^{-3}$    & $\,0.009$        & $1.63  \left(-2\right) $      & $1.11 \left(-2\right) $     & $8.85 \left(-3\right) $  & $0.30 $      \\[1ex]
    \end{tabular}
\end{ruledtabular}
\end{table*}

\section{Numerical Results And Discussion} \label{sec:consistency}

There are some reasonable assumptions upon which our probabilistic, physics-motivated approach relies, which we need to recall here. First, we assume that the test charge has a mass much larger than the electron effective mass  $m^{\ast}$, a common assumption in the computation of the electron lifetimes in semiconductors~\cite{caruso2016, marchetti2018}. Second, we consider the electron-test-charge scattering as an elastic process, similar to the assumption made for electron-ionized impurity collisions contributing to mobility in semiconductors~\cite{pines1966, chattopadhyay1981}. Furthermore, electrons at the Fermi surface scatter with collision energy equal to   $E_{\rm F} $, ranging from a few  \SI{}{\meV}  to about  \SI{20}{\eV} for the problem at hand.

\begin{figure}

\resizebox{0.50\textwidth}{!}{%
  \includegraphics{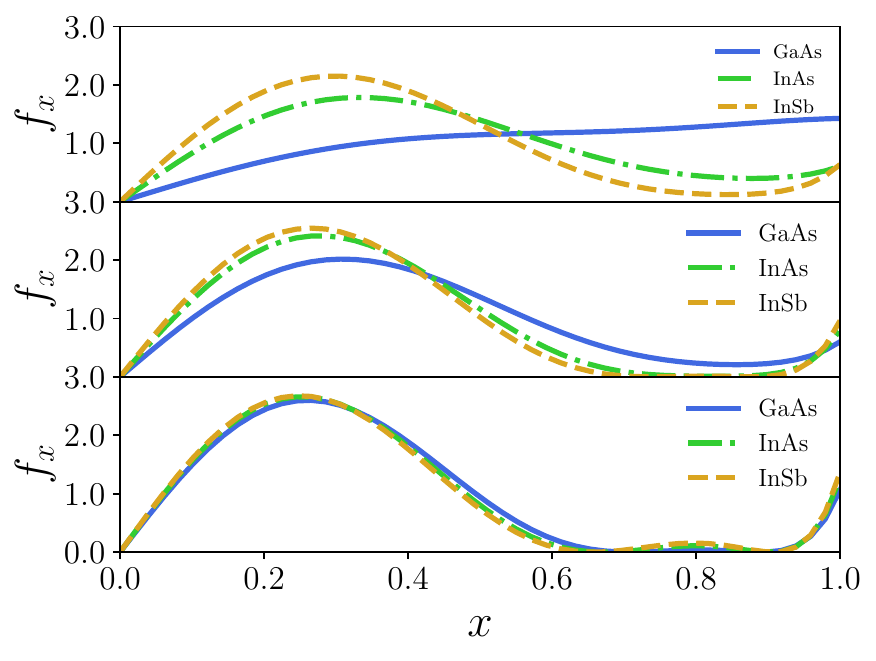}
}

\caption{The probability density function $f_x$ as function of the variable $x$ for GaAs, InAs and InSb. The panels (a), (b) and (c) correspond to the electron densities  \SI{}{ 10^{16}\cm^{-3}}, \SI{}{ 10^{18}\cm^{-3}} and \SI{}{10^{21}\cm^{-3}}, respectively.}
\label{fig:figure3}       
\end{figure}

In panels (a), (b), (c) of Fig.~\ref{fig:figure3} we plot the probability density function $f_x$ as function of the variable $x$ for GaAs, InAs and InSb doped with electron density \SI{}{ 10^{16}\cm^{-3}}, \SI{}{ 10^{18}\cm^{-3}} and \SI{}{10^{21}\cm^{-3}}, respectively.
The probability density functions  are calculated through Eq.~\ref{eq:pdf5}, and the respective values of $\delta_0, \delta_1, \delta_2$ on which each PDF is built on, are shown in Table~\ref{table:table2}.

With the exception of the PDF relative to GaAs at low electron density, as discussed below, the curves exhibit two common features: a maximum and a tail. Their maxima are reached for small values of $x$ ($x<0.4$), while their tails occur at large values of  $x\approx 1$. The latter finding implies that at zero temperature, there exists a non-negligible probability for back-scattering corresponding to the momentum transfer $q= 2 k_{\rm F}$. Note that their respective expectation value $\Bar{x}$ decreases as the electron density increases approximately reaching the value $0.3$ at the highest electron density (as reported in Table~\ref{table:table2}). This fact, together with the emerging similar PDF profile, irrespective of the type of material, is a direct consequence of the smallness of the respective phase shifts $\delta_l \sim 10^{-2}$. Thus, our analysis confirms that the Thomas-Fermi model is a high electron density approximation, and so it is certainly valid at metallic regimes.

The anomalous behavior of GaAs's probability density function, as seen in the respective curve (solid line) in panel (a) of Fig.~\ref{fig:figure3}, is a direct consequence of the lack of a maximum in the low dopant region, resulting in a linear trend. This linear trend leads to the emergence of a particularly large tail in the probability density function of such a material. Overall, one would expect a substantial average momentum transfer in this case. 
In fact, we found that  $\Bar{x}$ ranges from $0.59$ to $0.39$ at low and intermediate densities, respectively, as reported in Table~\ref{table:table2}. Therefore, the use of the Thomas-Fermi potential becomes highly questionable in such dopant concentrations.

To understand the physical reason of the  anomalous trend of the probability density function for GaAs, one needs to recall that the phase shifts crucially depend on the details of the screening as  observed by Bethe and Salpeter~\cite{bethe1957}. First, we note that in such a case, the magnitude of the computed  $\delta_0, \delta_1$ is the largest among all, as shown in the first row of Table~\ref{table:table2}. Second, we observe that $\delta_0 \approx 1.36$ when  \SI{}{ 10^{16}\cm^{-3}}, thereby compromising the Born approximation, which is valid when each $\delta_l \ll 1$ (or equivalently $\delta_l \ll \pi/2$)~\cite{bethe1957}. The latter fact is indeed a clear signature of a strong electron scattering. There are two reasons for this occurrence.  First, the Thomas-Fermi potential's range $\sim 1/q_{\rm TF}$ increases  as dopant concentrations decrease. Second,  the collision energy, i.e., $E_{\rm F} \sim n^{\frac{2}{3}}$, also decreases as $n$ decreases. Moreover, it is dramatically lowered for GaAs due to its largest effective mass $m^{\ast}$ among the materials under scrutiny (see Table~\ref{table:table1}).

\begin{figure}
\resizebox{0.50\textwidth}{!}{%
  \includegraphics{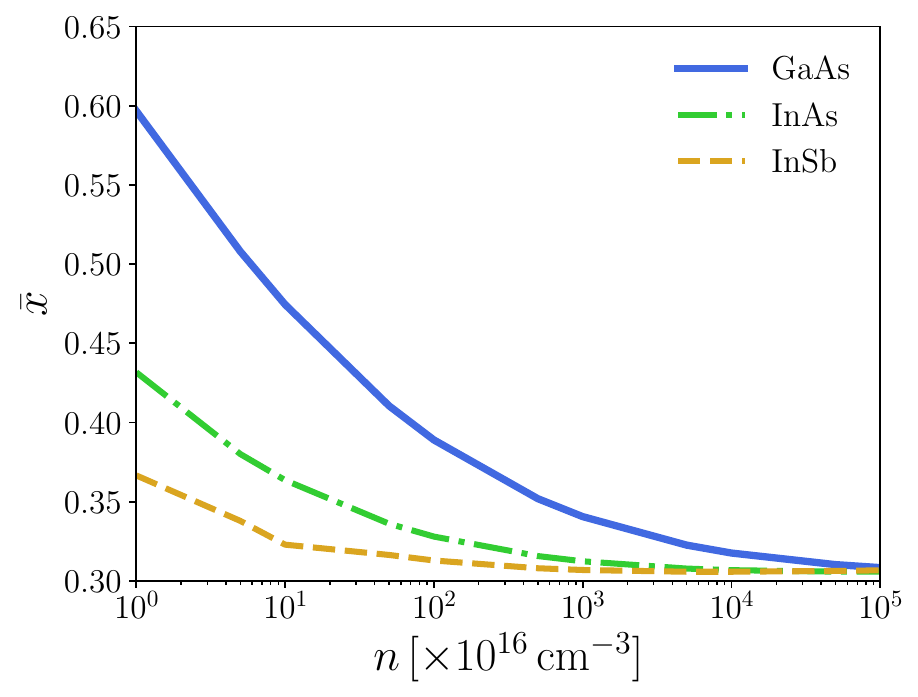}
}

\caption{The expectation value $\Bar{x}$ of the variable $x$ as function of the electron density $n$ for GaAs, InAs and InSb. }
\label{fig:figure4}       
\end{figure}

The expectation value $\Bar{x}$ as function of $n$ for densities ranging from \SI{}{ 10^{16}\cm^{-3}} to  \SI{}{ 10^{21}\cm^{-3}} for GaAs, InAs and InSb is shown in Fig.~\ref{fig:figure4}. Note that for each computation, we ascertained the probability density function is nonnegative everywhere. We found that this would be no the case if we had include additional contributions to PDF, for example, from the quantum phase shift $\delta_3$.  In Fig.~\ref{fig:figure4} all the curves exhibit a monotonically decreasing trend, converging to the value $\Bar{x} \approx 0.3$, irrespective of the material under scrutiny. Such a trend further confirms that the TF potential can be successfully employed in metallic regime. Nevertheless, due to the observed large values of $\Bar{x}$ for GaAs across most electron densities, using the TF potential for computing observables at zero temperature in such a material would likely yield inaccurate results.

We conclude by noting that the above analysis crucially depends on accurately numerically integrating the Riccati differential equation  (see Eq.~\ref{eq:phase}) in order to prevent possible inaccuracies and round-off errors~\cite{Romualdi2021}.

\section{Conclusion}

We revisited the theory of Thomas-Fermi potential for three-dimensional condensed matter systems according to the linear response theory and random phase approximation. Subsequently, we proposed a rigorous probabilistic method for assessing the validity of the Thomas-Fermi approximation. This method should ensure the correct computation of observables of interest for a given material. Although our focus has been on the fermionic dynamics on the Fermi surface at  zero temperature, this approach can be readily extended to the case of finite temperature in three-dimensional systems. To achieve this, researchers need to provide the appropriate temperature-dependence for both the momentum transfer and Thomas-Fermi's screening parameter~\cite{debye1923, chattopadhyay1981, nair2024induced}.

Finally, it is worth noting that our method, which relies on the variable phase method for accurately computing quantum phase shifts, can also be extended to apply the Thomas-Fermi potential to low-dimensional systems~\cite{portnoi1997}, including materials like graphene~\cite{geim2007, bolotin2008, morozov2008}. In the latter case, the proposed method becomes feasible due to the variable phase method's capability to handle Dirac (or equivalently, relativistic) particles~\cite{calogero1967, stone2012}.

\begin{acknowledgments}
The author thanks Dr. Bengt Eliasson for correspondence regarding the exponential cosine screened Coulomb potential and Dr. Alexandros Karam for obtaining the expansions in Eqs.~\ref{eq:rpa4},  \ref{eq:rpa5} through Wolfram Mathematica~\cite{ram2010}. The author also takes this opportunity to express his gratitude to Dr. Dorothea Golze, Dr. Mark Dvorak, and Prof. Patrick Rinke for giving him the opportunity to present preliminary results at the GW goes large-scale (GW-XL) Workshop in Helsinki.
\end{acknowledgments}

\bibliography{references}{}

%merlin.mbs apsrev4-1.bst 2010-07-25 4.21a (PWD, AO, DPC) hacked
%Control: key (0)
%Control: author (72) initials jnrlst
%Control: editor formatted (1) identically to author
%Control: production of article title (-1) disabled
%Control: page (0) single
%Control: year (1) truncated
%Control: production of eprint (0) enabled
\begin{thebibliography}{59}%
\makeatletter
\providecommand \@ifxundefined [1]{%
 \@ifx{#1\undefined}
}%
\providecommand \@ifnum [1]{%
 \ifnum #1\expandafter \@firstoftwo
 \else \expandafter \@secondoftwo
 \fi
}%
\providecommand \@ifx [1]{%
 \ifx #1\expandafter \@firstoftwo
 \else \expandafter \@secondoftwo
 \fi
}%
\providecommand \natexlab [1]{#1}%
\providecommand \enquote  [1]{``#1''}%
\providecommand \bibnamefont  [1]{#1}%
\providecommand \bibfnamefont [1]{#1}%
\providecommand \citenamefont [1]{#1}%
\providecommand \href@noop [0]{\@secondoftwo}%
\providecommand \href [0]{\begingroup \@sanitize@url \@href}%
\providecommand \@href[1]{\@@startlink{#1}\@@href}%
\providecommand \@@href[1]{\endgroup#1\@@endlink}%
\providecommand \@sanitize@url [0]{\catcode `\\12\catcode `\$12\catcode
  `\&12\catcode `\#12\catcode `\^12\catcode `\_12\catcode `\%12\relax}%
\providecommand \@@startlink[1]{}%
\providecommand \@@endlink[0]{}%
\providecommand \url  [0]{\begingroup\@sanitize@url \@url }%
\providecommand \@url [1]{\endgroup\@href {#1}{\urlprefix }}%
\providecommand \urlprefix  [0]{URL }%
\providecommand \Eprint [0]{\href }%
\providecommand \doibase [0]{http://dx.doi.org/}%
\providecommand \selectlanguage [0]{\@gobble}%
\providecommand \bibinfo  [0]{\@secondoftwo}%
\providecommand \bibfield  [0]{\@secondoftwo}%
\providecommand \translation [1]{[#1]}%
\providecommand \BibitemOpen [0]{}%
\providecommand \bibitemStop [0]{}%
\providecommand \bibitemNoStop [0]{.\EOS\space}%
\providecommand \EOS [0]{\spacefactor3000\relax}%
\providecommand \BibitemShut  [1]{\csname bibitem#1\endcsname}%
\let\auto@bib@innerbib\@empty
%</preamble>
\bibitem [{\citenamefont {Bohm}\ and\ \citenamefont {Pines}(1953)}]{bohm1953}%
  \BibitemOpen
  \bibfield  {author} {\bibinfo {author} {\bibfnamefont {D.}~\bibnamefont
  {Bohm}}\ and\ \bibinfo {author} {\bibfnamefont {D.}~\bibnamefont {Pines}},\
  }\href {\doibase 10.1103/PhysRev.92.609} {\bibfield  {journal} {\bibinfo
  {journal} {Phys. Rev.}\ }\textbf {\bibinfo {volume} {92}},\ \bibinfo {pages}
  {609} (\bibinfo {year} {1953})}\BibitemShut {NoStop}%
\bibitem [{\citenamefont {Pines}(2016)}]{pines2016}%
  \BibitemOpen
  \bibfield  {author} {\bibinfo {author} {\bibfnamefont {D.}~\bibnamefont
  {Pines}},\ }\href {\doibase 10.1088/0034-4885/79/9/092501} {\bibfield
  {journal} {\bibinfo  {journal} {Reports on Progress in Physics}\ }\textbf
  {\bibinfo {volume} {79}},\ \bibinfo {pages} {092501} (\bibinfo {year}
  {2016})}\BibitemShut {NoStop}%
\bibitem [{\citenamefont {Ren}\ \emph {et~al.}(2012)\citenamefont {Ren},
  \citenamefont {Rinke}, \citenamefont {Joas},\ and\ \citenamefont
  {Scheffler}}]{ren2012}%
  \BibitemOpen
  \bibfield  {author} {\bibinfo {author} {\bibfnamefont {X.}~\bibnamefont
  {Ren}}, \bibinfo {author} {\bibfnamefont {P.}~\bibnamefont {Rinke}}, \bibinfo
  {author} {\bibfnamefont {C.}~\bibnamefont {Joas}}, \ and\ \bibinfo {author}
  {\bibfnamefont {M.}~\bibnamefont {Scheffler}},\ }\href@noop {} {\bibfield
  {journal} {\bibinfo  {journal} {J. Mater. Sci.}\ }\textbf {\bibinfo {volume}
  {47}},\ \bibinfo {pages} {7447} (\bibinfo {year} {2012})}\BibitemShut
  {NoStop}%
\bibitem [{\citenamefont {Penn}(1987)}]{penn1987}%
  \BibitemOpen
  \bibfield  {author} {\bibinfo {author} {\bibfnamefont {D.~R.}\ \bibnamefont
  {Penn}},\ }\href@noop {} {\bibfield  {journal} {\bibinfo  {journal} {Phys.
  Rev. B}\ }\textbf {\bibinfo {volume} {35}},\ \bibinfo {pages} {482} (\bibinfo
  {year} {1987})}\BibitemShut {NoStop}%
\bibitem [{\citenamefont {Emfietzoglou}\ \emph {et~al.}(2013)\citenamefont
  {Emfietzoglou}, \citenamefont {Kyriakou}, \citenamefont {Garcia-Molina},\
  and\ \citenamefont {Abril}}]{enfietzoglou2013}%
  \BibitemOpen
  \bibfield  {author} {\bibinfo {author} {\bibfnamefont {D.}~\bibnamefont
  {Emfietzoglou}}, \bibinfo {author} {\bibfnamefont {I.}~\bibnamefont
  {Kyriakou}}, \bibinfo {author} {\bibfnamefont {R.}~\bibnamefont
  {Garcia-Molina}}, \ and\ \bibinfo {author} {\bibfnamefont {I.}~\bibnamefont
  {Abril}},\ }\href@noop {} {\bibfield  {journal} {\bibinfo  {journal} {Journal
  of Applied Physics}\ }\textbf {\bibinfo {volume} {114}},\ \bibinfo {pages}
  {144907} (\bibinfo {year} {2013})}\BibitemShut {NoStop}%
\bibitem [{\citenamefont {Hedin}(1965)}]{hedin1965}%
  \BibitemOpen
  \bibfield  {author} {\bibinfo {author} {\bibfnamefont {L.}~\bibnamefont
  {Hedin}},\ }\href@noop {} {\bibfield  {journal} {\bibinfo  {journal} {Phys.
  Rev.}\ }\textbf {\bibinfo {volume} {139}},\ \bibinfo {pages} {A796} (\bibinfo
  {year} {1965})}\BibitemShut {NoStop}%
\bibitem [{\citenamefont {Golze}\ \emph {et~al.}(2019)\citenamefont {Golze},
  \citenamefont {Dvorak},\ and\ \citenamefont {Rinke}}]{golze2019}%
  \BibitemOpen
  \bibfield  {author} {\bibinfo {author} {\bibfnamefont {D.}~\bibnamefont
  {Golze}}, \bibinfo {author} {\bibfnamefont {M.}~\bibnamefont {Dvorak}}, \
  and\ \bibinfo {author} {\bibfnamefont {P.}~\bibnamefont {Rinke}},\ }\href
  {\doibase 10.3389/fchem.2019.00377} {\bibfield  {journal} {\bibinfo
  {journal} {Frontiers in Chemistry}\ }\textbf {\bibinfo {volume} {7}},\
  \bibinfo {pages} {377} (\bibinfo {year} {2019})}\BibitemShut {NoStop}%
\bibitem [{\citenamefont {Marzari}\ \emph {et~al.}(2021)\citenamefont
  {Marzari}, \citenamefont {Ferretti},\ and\ \citenamefont
  {Wolverton}}]{Marzari2021}%
  \BibitemOpen
  \bibfield  {author} {\bibinfo {author} {\bibfnamefont {N.}~\bibnamefont
  {Marzari}}, \bibinfo {author} {\bibfnamefont {A.}~\bibnamefont {Ferretti}}, \
  and\ \bibinfo {author} {\bibfnamefont {C.}~\bibnamefont {Wolverton}},\ }\href
  {\doibase 10.1038/s41563-021-01013-3} {\bibfield  {journal} {\bibinfo
  {journal} {Nature Materials}\ }\textbf {\bibinfo {volume} {20}},\ \bibinfo
  {pages} {736} (\bibinfo {year} {2021})}\BibitemShut {NoStop}%
\bibitem [{\citenamefont {Ashcroft}\ and\ \citenamefont
  {Mermin}(1976)}]{ashcroft1976}%
  \BibitemOpen
  \bibfield  {author} {\bibinfo {author} {\bibfnamefont {N.~W.}\ \bibnamefont
  {Ashcroft}}\ and\ \bibinfo {author} {\bibfnamefont {M.~D.}\ \bibnamefont
  {Mermin}},\ }\href@noop {} {\emph {\bibinfo {title} {{Solid State
  Physics}}}}\ (\bibinfo  {publisher} {Saunders College},\ \bibinfo {address}
  {Philadelphia},\ \bibinfo {year} {1976})\BibitemShut {NoStop}%
\bibitem [{\citenamefont {Das~Sarma}\ and\ \citenamefont
  {Stern}(1985)}]{dasSarma1985}%
  \BibitemOpen
  \bibfield  {author} {\bibinfo {author} {\bibfnamefont {S.}~\bibnamefont
  {Das~Sarma}}\ and\ \bibinfo {author} {\bibfnamefont {F.}~\bibnamefont
  {Stern}},\ }\href {\doibase 10.1103/PhysRevB.32.8442} {\bibfield  {journal}
  {\bibinfo  {journal} {Phys. Rev. B}\ }\textbf {\bibinfo {volume} {32}},\
  \bibinfo {pages} {8442} (\bibinfo {year} {1985})}\BibitemShut {NoStop}%
\bibitem [{\citenamefont {Caruso}\ and\ \citenamefont
  {Giustino}(2016)}]{caruso2016}%
  \BibitemOpen
  \bibfield  {author} {\bibinfo {author} {\bibfnamefont {F.}~\bibnamefont
  {Caruso}}\ and\ \bibinfo {author} {\bibfnamefont {F.}~\bibnamefont
  {Giustino}},\ }\href@noop {} {\bibfield  {journal} {\bibinfo  {journal}
  {Phys. Rev. B}\ }\textbf {\bibinfo {volume} {94}},\ \bibinfo {pages} {115208}
  (\bibinfo {year} {2016})}\BibitemShut {NoStop}%
\bibitem [{\citenamefont {Hall}(1962)}]{hall1962}%
  \BibitemOpen
  \bibfield  {author} {\bibinfo {author} {\bibfnamefont {G.}~\bibnamefont
  {Hall}},\ }\href@noop {} {\bibfield  {journal} {\bibinfo  {journal} {Journal
  of Physics and Chemistry of Solids}\ }\textbf {\bibinfo {volume} {23}},\
  \bibinfo {pages} {1147} (\bibinfo {year} {1962})}\BibitemShut {NoStop}%
\bibitem [{\citenamefont {Chattopadhyay}\ and\ \citenamefont
  {Queisser}(1981)}]{chattopadhyay1981}%
  \BibitemOpen
  \bibfield  {author} {\bibinfo {author} {\bibfnamefont {D.}~\bibnamefont
  {Chattopadhyay}}\ and\ \bibinfo {author} {\bibfnamefont {H.~J.}\ \bibnamefont
  {Queisser}},\ }\href@noop {} {\bibfield  {journal} {\bibinfo  {journal} {Rev.
  Mod. Phys.}\ }\textbf {\bibinfo {volume} {53}},\ \bibinfo {pages} {745}
  (\bibinfo {year} {1981})}\BibitemShut {NoStop}%
\bibitem [{\citenamefont {Meyer}\ and\ \citenamefont
  {Bartoli}(1981{\natexlab{a}})}]{meyer1981}%
  \BibitemOpen
  \bibfield  {author} {\bibinfo {author} {\bibfnamefont {J.~R.}\ \bibnamefont
  {Meyer}}\ and\ \bibinfo {author} {\bibfnamefont {F.~J.}\ \bibnamefont
  {Bartoli}},\ }\href {\doibase 10.1103/PhysRevB.23.5413} {\bibfield  {journal}
  {\bibinfo  {journal} {Phys. Rev. B}\ }\textbf {\bibinfo {volume} {23}},\
  \bibinfo {pages} {5413} (\bibinfo {year} {1981}{\natexlab{a}})}\BibitemShut
  {NoStop}%
\bibitem [{\citenamefont {Meyer}\ and\ \citenamefont
  {Bartoli}(1981{\natexlab{b}})}]{meyer1981a}%
  \BibitemOpen
  \bibfield  {author} {\bibinfo {author} {\bibfnamefont {J.~R.}\ \bibnamefont
  {Meyer}}\ and\ \bibinfo {author} {\bibfnamefont {F.~J.}\ \bibnamefont
  {Bartoli}},\ }\href@noop {} {\bibfield  {journal} {\bibinfo  {journal} {Phys.
  Rev. B}\ }\textbf {\bibinfo {volume} {24}},\ \bibinfo {pages} {2089}
  (\bibinfo {year} {1981}{\natexlab{b}})}\BibitemShut {NoStop}%
\bibitem [{\citenamefont {Shepherd}\ and\ \citenamefont
  {Gr\"uneis}(2013)}]{shepherd2013}%
  \BibitemOpen
  \bibfield  {author} {\bibinfo {author} {\bibfnamefont {J.~J.}\ \bibnamefont
  {Shepherd}}\ and\ \bibinfo {author} {\bibfnamefont {A.}~\bibnamefont
  {Gr\"uneis}},\ }\href@noop {} {\bibfield  {journal} {\bibinfo  {journal}
  {Phys. Rev. Lett.}\ }\textbf {\bibinfo {volume} {110}},\ \bibinfo {pages}
  {226401} (\bibinfo {year} {2013})}\BibitemShut {NoStop}%
\bibitem [{\citenamefont {Hwang}\ and\ \citenamefont
  {Das~Sarma}(2007)}]{hwang2007}%
  \BibitemOpen
  \bibfield  {author} {\bibinfo {author} {\bibfnamefont {E.~H.}\ \bibnamefont
  {Hwang}}\ and\ \bibinfo {author} {\bibfnamefont {S.}~\bibnamefont
  {Das~Sarma}},\ }\href {\doibase 10.1103/PhysRevB.75.205418} {\bibfield
  {journal} {\bibinfo  {journal} {Phys. Rev. B}\ }\textbf {\bibinfo {volume}
  {75}},\ \bibinfo {pages} {205418} (\bibinfo {year} {2007})}\BibitemShut
  {NoStop}%
\bibitem [{\citenamefont {Takimoto}(1959)}]{takimoto1959}%
  \BibitemOpen
  \bibfield  {author} {\bibinfo {author} {\bibfnamefont {N.}~\bibnamefont
  {Takimoto}},\ }\href@noop {} {\bibfield  {journal} {\bibinfo  {journal}
  {Journal of the Physical Society of Japan}\ }\textbf {\bibinfo {volume}
  {14}},\ \bibinfo {pages} {1142} (\bibinfo {year} {1959})}\BibitemShut
  {NoStop}%
\bibitem [{\citenamefont {Bonch-Bruevich}\ and\ \citenamefont
  {Tyablikov}(1962)}]{bonch-bruevich1962}%
  \BibitemOpen
  \bibfield  {author} {\bibinfo {author} {\bibfnamefont {V.~L.}\ \bibnamefont
  {Bonch-Bruevich}}\ and\ \bibinfo {author} {\bibfnamefont {S.~V.}\
  \bibnamefont {Tyablikov}},\ }\href@noop {} {\emph {\bibinfo {title} {{Green
  Function Method in Statistical Mechanics}}}}\ (\bibinfo  {publisher}
  {North-Holland Publishing Company},\ \bibinfo {address} {Amsterdam},\
  \bibinfo {year} {1962})\BibitemShut {NoStop}%
\bibitem [{\citenamefont {Lam}\ and\ \citenamefont {Varshni}(1972)}]{lam1972}%
  \BibitemOpen
  \bibfield  {author} {\bibinfo {author} {\bibfnamefont {C.~S.}\ \bibnamefont
  {Lam}}\ and\ \bibinfo {author} {\bibfnamefont {Y.~P.}\ \bibnamefont
  {Varshni}},\ }\href {\doibase 10.1103/PhysRevA.6.1391} {\bibfield  {journal}
  {\bibinfo  {journal} {Phys. Rev. A}\ }\textbf {\bibinfo {volume} {6}},\
  \bibinfo {pages} {1391} (\bibinfo {year} {1972})}\BibitemShut {NoStop}%
\bibitem [{\citenamefont {Shukla}\ and\ \citenamefont
  {Eliasson}(2008)}]{shukla2008}%
  \BibitemOpen
  \bibfield  {author} {\bibinfo {author} {\bibfnamefont {P.~K.}\ \bibnamefont
  {Shukla}}\ and\ \bibinfo {author} {\bibfnamefont {B.}~\bibnamefont
  {Eliasson}},\ }\href@noop {} {\bibfield  {journal} {\bibinfo  {journal}
  {Physics Letters A}\ }\textbf {\bibinfo {volume} {372}},\ \bibinfo {pages}
  {2897 } (\bibinfo {year} {2008})}\BibitemShut {NoStop}%
\bibitem [{\citenamefont {Shukla}\ and\ \citenamefont
  {Eliasson}(2012)}]{shukla2012}%
  \BibitemOpen
  \bibfield  {author} {\bibinfo {author} {\bibfnamefont {P.~K.}\ \bibnamefont
  {Shukla}}\ and\ \bibinfo {author} {\bibfnamefont {B.}~\bibnamefont
  {Eliasson}},\ }\href {\doibase 10.1103/PhysRevLett.108.165007} {\bibfield
  {journal} {\bibinfo  {journal} {Phys. Rev. Lett.}\ }\textbf {\bibinfo
  {volume} {108}},\ \bibinfo {pages} {165007} (\bibinfo {year}
  {2012})}\BibitemShut {NoStop}%
\bibitem [{\citenamefont {Nayek}\ and\ \citenamefont
  {Ghoshal}(2012)}]{nayek2012}%
  \BibitemOpen
  \bibfield  {author} {\bibinfo {author} {\bibfnamefont {S.}~\bibnamefont
  {Nayek}}\ and\ \bibinfo {author} {\bibfnamefont {A.}~\bibnamefont
  {Ghoshal}},\ }\href@noop {} {\bibfield  {journal} {\bibinfo  {journal}
  {Physica Scripta}\ }\textbf {\bibinfo {volume} {85}},\ \bibinfo {pages}
  {035301} (\bibinfo {year} {2012})}\BibitemShut {NoStop}%
\bibitem [{\citenamefont {Qi}\ \emph {et~al.}(2016)\citenamefont {Qi},
  \citenamefont {Wang},\ and\ \citenamefont {Janev}}]{qi2016}%
  \BibitemOpen
  \bibfield  {author} {\bibinfo {author} {\bibfnamefont {Y.~Y.}\ \bibnamefont
  {Qi}}, \bibinfo {author} {\bibfnamefont {J.~G.}\ \bibnamefont {Wang}}, \ and\
  \bibinfo {author} {\bibfnamefont {R.~K.}\ \bibnamefont {Janev}},\ }\href@noop
  {} {\bibfield  {journal} {\bibinfo  {journal} {Physics of Plasmas}\ }\textbf
  {\bibinfo {volume} {23}},\ \bibinfo {pages} {073302} (\bibinfo {year}
  {2016})}\BibitemShut {NoStop}%
\bibitem [{\citenamefont {Janev}\ \emph {et~al.}(2016)\citenamefont {Janev},
  \citenamefont {Zhang},\ and\ \citenamefont {Wang}}]{janev2016}%
  \BibitemOpen
  \bibfield  {author} {\bibinfo {author} {\bibfnamefont {R.~K.}\ \bibnamefont
  {Janev}}, \bibinfo {author} {\bibfnamefont {S.}~\bibnamefont {Zhang}}, \ and\
  \bibinfo {author} {\bibfnamefont {J.}~\bibnamefont {Wang}},\ }\href@noop {}
  {\bibfield  {journal} {\bibinfo  {journal} {Matter and Radiation at
  Extremes}\ }\textbf {\bibinfo {volume} {1}},\ \bibinfo {pages} {237 }
  (\bibinfo {year} {2016})}\BibitemShut {NoStop}%
\bibitem [{\citenamefont {Munjal}\ \emph {et~al.}(2017)\citenamefont {Munjal},
  \citenamefont {Silotia},\ and\ \citenamefont {Prasad}}]{munjal2017}%
  \BibitemOpen
  \bibfield  {author} {\bibinfo {author} {\bibfnamefont {D.}~\bibnamefont
  {Munjal}}, \bibinfo {author} {\bibfnamefont {P.}~\bibnamefont {Silotia}}, \
  and\ \bibinfo {author} {\bibfnamefont {V.}~\bibnamefont {Prasad}},\
  }\href@noop {} {\bibfield  {journal} {\bibinfo  {journal} {Physics of
  Plasmas}\ }\textbf {\bibinfo {volume} {24}},\ \bibinfo {pages} {122118}
  (\bibinfo {year} {2017})}\BibitemShut {NoStop}%
\bibitem [{\citenamefont {Chen}\ \emph {et~al.}(2023)\citenamefont {Chen},
  \citenamefont {Zhao}, \citenamefont {Chen}, \citenamefont {Qi}, \citenamefont
  {Liu}, \citenamefont {Wu},\ and\ \citenamefont {Wang}}]{chen2023}%
  \BibitemOpen
  \bibfield  {author} {\bibinfo {author} {\bibfnamefont {C.}~\bibnamefont
  {Chen}}, \bibinfo {author} {\bibfnamefont {G.~P.}\ \bibnamefont {Zhao}},
  \bibinfo {author} {\bibfnamefont {Z.~B.}\ \bibnamefont {Chen}}, \bibinfo
  {author} {\bibfnamefont {Y.~Y.}\ \bibnamefont {Qi}}, \bibinfo {author}
  {\bibfnamefont {L.}~\bibnamefont {Liu}}, \bibinfo {author} {\bibfnamefont
  {Y.}~\bibnamefont {Wu}}, \ and\ \bibinfo {author} {\bibfnamefont {J.~G.}\
  \bibnamefont {Wang}},\ }\href@noop {} {\bibfield  {journal} {\bibinfo
  {journal} {Physics of Plasmas}\ }\textbf {\bibinfo {volume} {30}},\ \bibinfo
  {pages} {123503} (\bibinfo {year} {2023})}\BibitemShut {NoStop}%
\bibitem [{\citenamefont {Polchinski}(1999)}]{polchinski1999effective}%
  \BibitemOpen
  \bibfield  {author} {\bibinfo {author} {\bibfnamefont {J.}~\bibnamefont
  {Polchinski}},\ }\href@noop {} {\enquote {\bibinfo {title} {Effective field
  theory and the fermi surface},}\ } (\bibinfo {year} {1999}),\ \Eprint
  {http://arxiv.org/abs/hep-th/9210046} {arXiv:hep-th/9210046 [hep-th]}
  \BibitemShut {NoStop}%
\bibitem [{\citenamefont {Bethe}\ and\ \citenamefont
  {Salpeter}(1957)}]{bethe1957}%
  \BibitemOpen
  \bibfield  {author} {\bibinfo {author} {\bibfnamefont {H.~A.}\ \bibnamefont
  {Bethe}}\ and\ \bibinfo {author} {\bibfnamefont {E.~E.}\ \bibnamefont
  {Salpeter}},\ }\href@noop {} {\emph {\bibinfo {title} {{Quantum Mechanics of
  One- and Two-Electron Atoms}}}}\ (\bibinfo  {publisher} {Springer-Verlag},\
  \bibinfo {address} {Berlin G{\"o}ttingen Heidelberg},\ \bibinfo {year}
  {1957})\BibitemShut {NoStop}%
\bibitem [{\citenamefont {Calogero}(1963)}]{calogero1963}%
  \BibitemOpen
  \bibfield  {author} {\bibinfo {author} {\bibfnamefont {F.}~\bibnamefont
  {Calogero}},\ }\href@noop {} {\bibfield  {journal} {\bibinfo  {journal} {Il
  Nuovo Cimento}\ }\textbf {\bibinfo {volume} {27}},\ \bibinfo {pages} {261}
  (\bibinfo {year} {1963})}\BibitemShut {NoStop}%
\bibitem [{\citenamefont {Calogero}(1967)}]{calogero1967}%
  \BibitemOpen
  \bibfield  {author} {\bibinfo {author} {\bibfnamefont {F.}~\bibnamefont
  {Calogero}},\ }\href@noop {} {\emph {\bibinfo {title} {{ Variable Phase
  Approach to Potential Scattering }}}}\ (\bibinfo  {publisher} {Academic
  Press},\ \bibinfo {address} {New York and London},\ \bibinfo {year}
  {1967})\BibitemShut {NoStop}%
\bibitem [{\citenamefont {Babikov}(1967)}]{babikov1967}%
  \BibitemOpen
  \bibfield  {author} {\bibinfo {author} {\bibfnamefont {V.~V.}\ \bibnamefont
  {Babikov}},\ }\href@noop {} {\bibfield  {journal} {\bibinfo  {journal}
  {Soviet Physics Uspekhi}\ }\textbf {\bibinfo {volume} {10}},\ \bibinfo
  {pages} {271} (\bibinfo {year} {1967})}\BibitemShut {NoStop}%
\bibitem [{\citenamefont {Madelung}(1991)}]{madelung1991}%
  \BibitemOpen
  \bibfield  {author} {\bibinfo {author} {\bibfnamefont {O.}~\bibnamefont
  {Madelung}},\ }\href@noop {} {\emph {\bibinfo {title} {{Semiconductors Group
  IV Elements and III-V Compounds }}}}\ (\bibinfo  {publisher}
  {Springer-Verlag},\ \bibinfo {address} {Berlin Heidelberg, Germany},\
  \bibinfo {year} {1991})\BibitemShut {NoStop}%
\bibitem [{\citenamefont {Vurgaftman}\ \emph {et~al.}(2001)\citenamefont
  {Vurgaftman}, \citenamefont {Meyer},\ and\ \citenamefont
  {Ram-Mohan}}]{vurgaftman2001}%
  \BibitemOpen
  \bibfield  {author} {\bibinfo {author} {\bibfnamefont {I.}~\bibnamefont
  {Vurgaftman}}, \bibinfo {author} {\bibfnamefont {J.~R.}\ \bibnamefont
  {Meyer}}, \ and\ \bibinfo {author} {\bibfnamefont {L.~R.}\ \bibnamefont
  {Ram-Mohan}},\ }\href {\doibase 10.1063/1.1368156} {\bibfield  {journal}
  {\bibinfo  {journal} {Journal of Applied Physics}\ }\textbf {\bibinfo
  {volume} {89}},\ \bibinfo {pages} {5815} (\bibinfo {year}
  {2001})}\BibitemShut {NoStop}%
\bibitem [{\citenamefont {Giuliani}\ and\ \citenamefont
  {Vignale}(2005)}]{giuliani2005}%
  \BibitemOpen
  \bibfield  {author} {\bibinfo {author} {\bibfnamefont {G.}~\bibnamefont
  {Giuliani}}\ and\ \bibinfo {author} {\bibfnamefont {G.}~\bibnamefont
  {Vignale}},\ }\href@noop {} {\emph {\bibinfo {title} {{Quantum Theory of
  Electron Liquid}}}}\ (\bibinfo  {publisher} {Cambridge University Press},\
  \bibinfo {address} {Cambridge, UK},\ \bibinfo {year} {2005})\BibitemShut
  {NoStop}%
\bibitem [{\citenamefont {Schliemann}(2011)}]{schliemann2011}%
  \BibitemOpen
  \bibfield  {author} {\bibinfo {author} {\bibfnamefont {J.}~\bibnamefont
  {Schliemann}},\ }\href@noop {} {\bibfield  {journal} {\bibinfo  {journal}
  {Phys. Rev. B}\ }\textbf {\bibinfo {volume} {84}},\ \bibinfo {pages} {155201}
  (\bibinfo {year} {2011})}\BibitemShut {NoStop}%
\bibitem [{\citenamefont {Lindhard}(1954)}]{lindhard1954}%
  \BibitemOpen
  \bibfield  {author} {\bibinfo {author} {\bibfnamefont {J.}~\bibnamefont
  {Lindhard}},\ }\href@noop {} {\bibfield  {journal} {\bibinfo  {journal} {Det
  Klg. Danske Vid. Selskab. Matematisk-fysiske Meddeleiser}\ }\textbf {\bibinfo
  {volume} {28}},\ \bibinfo {pages} {1} (\bibinfo {year} {1954})}\BibitemShut
  {NoStop}%
\bibitem [{\citenamefont {Gell-Mann}\ and\ \citenamefont
  {Brueckner}(1957)}]{gellmann1957}%
  \BibitemOpen
  \bibfield  {author} {\bibinfo {author} {\bibfnamefont {M.}~\bibnamefont
  {Gell-Mann}}\ and\ \bibinfo {author} {\bibfnamefont {K.~A.}\ \bibnamefont
  {Brueckner}},\ }\href@noop {} {\bibfield  {journal} {\bibinfo  {journal}
  {Phys. Rev.}\ }\textbf {\bibinfo {volume} {106}},\ \bibinfo {pages} {364}
  (\bibinfo {year} {1957})}\BibitemShut {NoStop}%
\bibitem [{\citenamefont {Adam}\ \emph {et~al.}(2007)\citenamefont {Adam},
  \citenamefont {Hwang}, \citenamefont {Galitski},\ and\ \citenamefont
  {Das~Sarma}}]{adam2007}%
  \BibitemOpen
  \bibfield  {author} {\bibinfo {author} {\bibfnamefont {S.}~\bibnamefont
  {Adam}}, \bibinfo {author} {\bibfnamefont {E.~H.}\ \bibnamefont {Hwang}},
  \bibinfo {author} {\bibfnamefont {V.~M.}\ \bibnamefont {Galitski}}, \ and\
  \bibinfo {author} {\bibfnamefont {S.}~\bibnamefont {Das~Sarma}},\ }\href@noop
  {} {\bibfield  {journal} {\bibinfo  {journal} {Proceedings of the National
  Academy of Sciences}\ }\textbf {\bibinfo {volume} {104}},\ \bibinfo {pages}
  {18392} (\bibinfo {year} {2007})}\BibitemShut {NoStop}%
\bibitem [{\citenamefont {Langer}\ and\ \citenamefont
  {Vosko}(1960)}]{langer1960}%
  \BibitemOpen
  \bibfield  {author} {\bibinfo {author} {\bibfnamefont {J.~S.}\ \bibnamefont
  {Langer}}\ and\ \bibinfo {author} {\bibfnamefont {S.~H.}\ \bibnamefont
  {Vosko}},\ }\href@noop {} {\bibfield  {journal} {\bibinfo  {journal} {Journal
  of Physics and Chemistry of Solids}\ }\textbf {\bibinfo {volume} {12}},\
  \bibinfo {pages} {196 } (\bibinfo {year} {1960})}\BibitemShut {NoStop}%
\bibitem [{\citenamefont {C.~McIrvine}(1960)}]{mcIrvine1960}%
  \BibitemOpen
  \bibfield  {author} {\bibinfo {author} {\bibfnamefont {E.}~\bibnamefont
  {C.~McIrvine}},\ }\href@noop {} {\bibfield  {journal} {\bibinfo  {journal}
  {Journal of the Physical Society of Japan}\ }\textbf {\bibinfo {volume}
  {15}},\ \bibinfo {pages} {928} (\bibinfo {year} {1960})}\BibitemShut
  {NoStop}%
\bibitem [{\citenamefont {Hohenberg}\ and\ \citenamefont
  {Kohn}(1964)}]{hohenberg1964}%
  \BibitemOpen
  \bibfield  {author} {\bibinfo {author} {\bibfnamefont {P.}~\bibnamefont
  {Hohenberg}}\ and\ \bibinfo {author} {\bibfnamefont {W.}~\bibnamefont
  {Kohn}},\ }\href {\doibase 10.1103/PhysRev.136.B864} {\bibfield  {journal}
  {\bibinfo  {journal} {Phys. Rev.}\ }\textbf {\bibinfo {volume} {136}},\
  \bibinfo {pages} {B864} (\bibinfo {year} {1964})}\BibitemShut {NoStop}%
\bibitem [{\citenamefont {Friedel}(1952)}]{friedel1952}%
  \BibitemOpen
  \bibfield  {author} {\bibinfo {author} {\bibfnamefont {J.}~\bibnamefont
  {Friedel}},\ }\href@noop {} {\bibfield  {journal} {\bibinfo  {journal} {Phil.
  Mag.}\ }\textbf {\bibinfo {volume} {43}} (\bibinfo {year}
  {1952})}\BibitemShut {NoStop}%
\bibitem [{\citenamefont {Stefanucci}\ and\ \citenamefont {van
  Leeuwen}(2013)}]{stefanucci2013}%
  \BibitemOpen
  \bibfield  {author} {\bibinfo {author} {\bibfnamefont {G.}~\bibnamefont
  {Stefanucci}}\ and\ \bibinfo {author} {\bibfnamefont {R.}~\bibnamefont {van
  Leeuwen}},\ }\href@noop {} {\emph {\bibinfo {title} {{Nonequilibrium
  Many-Body Theory of Quantum Systems A Modern Introduction}}}}\ (\bibinfo
  {publisher} {Cambridge University Press},\ \bibinfo {address} {Cambridge},\
  \bibinfo {year} {2013})\BibitemShut {NoStop}%
\bibitem [{\citenamefont {Schiff}(1968)}]{schiff1968}%
  \BibitemOpen
  \bibfield  {author} {\bibinfo {author} {\bibfnamefont {L.~I.}\ \bibnamefont
  {Schiff}},\ }\href@noop {} {\emph {\bibinfo {title} {{Quantum Mechanics}}}}\
  (\bibinfo  {publisher} {McGraw-Hill},\ \bibinfo {address} {Singapore},\
  \bibinfo {year} {1968})\BibitemShut {NoStop}%
\bibitem [{\citenamefont {Lundstrom}(2000)}]{lundstrom2000}%
  \BibitemOpen
  \bibfield  {author} {\bibinfo {author} {\bibfnamefont {M.}~\bibnamefont
  {Lundstrom}},\ }\href@noop {} {\emph {\bibinfo {title} {{ Fundamentals of
  Carrier Transport }}}}\ (\bibinfo  {publisher} {Cambridge University Press},\
  \bibinfo {address} {Cambridge},\ \bibinfo {year} {2000})\BibitemShut
  {NoStop}%
\bibitem [{\citenamefont {Marchetti}(2019)}]{marchetti2019}%
  \BibitemOpen
  \bibfield  {author} {\bibinfo {author} {\bibfnamefont {G.}~\bibnamefont
  {Marchetti}},\ }\href {\doibase 10.1063/1.5081631} {\bibfield  {journal}
  {\bibinfo  {journal} {Journal of Applied Physics}\ }\textbf {\bibinfo
  {volume} {126}},\ \bibinfo {pages} {045713} (\bibinfo {year} {2019})},\
  \Eprint {http://arxiv.org/abs/https://doi.org/10.1063/1.5081631}
  {https://doi.org/10.1063/1.5081631} \BibitemShut {NoStop}%
\bibitem [{\citenamefont {Deisenroth}\ \emph {et~al.}(2020)\citenamefont
  {Deisenroth}, \citenamefont {Faisal},\ and\ \citenamefont
  {Ong}}]{Deisenroth2020}%
  \BibitemOpen
  \bibfield  {author} {\bibinfo {author} {\bibfnamefont {M.~P.}\ \bibnamefont
  {Deisenroth}}, \bibinfo {author} {\bibfnamefont {A.~A.}\ \bibnamefont
  {Faisal}}, \ and\ \bibinfo {author} {\bibfnamefont {C.~S.}\ \bibnamefont
  {Ong}},\ }\href@noop {} {\emph {\bibinfo {title} {{ Mathematics for Machine
  Learning }}}}\ (\bibinfo  {publisher} {Cambridge University Press},\ \bibinfo
  {year} {2020})\BibitemShut {NoStop}%
\bibitem [{\citenamefont {Marchetti}(2018)}]{marchetti2018}%
  \BibitemOpen
  \bibfield  {author} {\bibinfo {author} {\bibfnamefont {G.}~\bibnamefont
  {Marchetti}},\ }\href {http://stacks.iop.org/0953-8984/30/i=47/a=475701}
  {\bibfield  {journal} {\bibinfo  {journal} {Journal of Physics: Condensed
  Matter}\ }\textbf {\bibinfo {volume} {30}},\ \bibinfo {pages} {475701}
  (\bibinfo {year} {2018})}\BibitemShut {NoStop}%
\bibitem [{\citenamefont {Pines}\ and\ \citenamefont
  {Nozieres}(1966)}]{pines1966}%
  \BibitemOpen
  \bibfield  {author} {\bibinfo {author} {\bibfnamefont {D.}~\bibnamefont
  {Pines}}\ and\ \bibinfo {author} {\bibfnamefont {P.}~\bibnamefont
  {Nozieres}},\ }\href@noop {} {\emph {\bibinfo {title} {{The Theory of Quantum
  Liquids,1:Normal Fermi Liquids}}}}\ (\bibinfo  {publisher} {W. A. Benjamin
  Inc.},\ \bibinfo {address} {New York},\ \bibinfo {year} {1966})\ p.\ \bibinfo
  {pages} {188}\BibitemShut {NoStop}%
\bibitem [{\citenamefont {Romualdi}\ and\ \citenamefont
  {Marchetti}(2021)}]{Romualdi2021}%
  \BibitemOpen
  \bibfield  {author} {\bibinfo {author} {\bibfnamefont {A.}~\bibnamefont
  {Romualdi}}\ and\ \bibinfo {author} {\bibfnamefont {G.}~\bibnamefont
  {Marchetti}},\ }\href {\doibase 10.1140/epjb/s10051-021-00261-1} {\bibfield
  {journal} {\bibinfo  {journal} {The European Physical Journal B}\ }\textbf
  {\bibinfo {volume} {94}},\ \bibinfo {pages} {249} (\bibinfo {year}
  {2021})}\BibitemShut {NoStop}%
\bibitem [{\citenamefont {Debye}\ and\ \citenamefont
  {H{\"u}ckel}(1923)}]{debye1923}%
  \BibitemOpen
  \bibfield  {author} {\bibinfo {author} {\bibfnamefont {P.}~\bibnamefont
  {Debye}}\ and\ \bibinfo {author} {\bibfnamefont {E.}~\bibnamefont
  {H{\"u}ckel}},\ }\href@noop {} {\bibfield  {journal} {\bibinfo  {journal}
  {Physikalische Zeitschrift}\ }\textbf {\bibinfo {volume} {24}},\ \bibinfo
  {pages} {185} (\bibinfo {year} {1923})}\BibitemShut {NoStop}%
\bibitem [{\citenamefont {Nair}\ \emph {et~al.}(2024)\citenamefont {Nair},
  \citenamefont {Pireddu},\ and\ \citenamefont {Rotenberg}}]{nair2024induced}%
  \BibitemOpen
  \bibfield  {author} {\bibinfo {author} {\bibfnamefont {S.}~\bibnamefont
  {Nair}}, \bibinfo {author} {\bibfnamefont {G.}~\bibnamefont {Pireddu}}, \
  and\ \bibinfo {author} {\bibfnamefont {B.}~\bibnamefont {Rotenberg}},\
  }\href@noop {} {\enquote {\bibinfo {title} {Induced charges in a thomas-fermi
  metal: insights from molecular simulations},}\ } (\bibinfo {year} {2024}),\
  \Eprint {http://arxiv.org/abs/2403.04487} {arXiv:2403.04487
  [physics.comp-ph]} \BibitemShut {NoStop}%
\bibitem [{\citenamefont {Portnoi}\ and\ \citenamefont
  {Galbraith}(1997)}]{portnoi1997}%
  \BibitemOpen
  \bibfield  {author} {\bibinfo {author} {\bibfnamefont {M.}~\bibnamefont
  {Portnoi}}\ and\ \bibinfo {author} {\bibfnamefont {I.}~\bibnamefont
  {Galbraith}},\ }\href {\doibase
  https://doi.org/10.1016/S0038-1098(97)00203-2} {\bibfield  {journal}
  {\bibinfo  {journal} {Solid State Communications}\ }\textbf {\bibinfo
  {volume} {103}},\ \bibinfo {pages} {325} (\bibinfo {year}
  {1997})}\BibitemShut {NoStop}%
\bibitem [{\citenamefont {Geim}\ and\ \citenamefont
  {Novoselov}(2007)}]{geim2007}%
  \BibitemOpen
  \bibfield  {author} {\bibinfo {author} {\bibfnamefont {A.}~\bibnamefont
  {Geim}}\ and\ \bibinfo {author} {\bibfnamefont {K.}~\bibnamefont
  {Novoselov}},\ }\href@noop {} {\bibfield  {journal} {\bibinfo  {journal}
  {Nature Materials}\ }\textbf {\bibinfo {volume} {6}},\ \bibinfo {pages} {183}
  (\bibinfo {year} {2007})}\BibitemShut {NoStop}%
\bibitem [{\citenamefont {Bolotin}\ \emph {et~al.}(2008)\citenamefont
  {Bolotin}, \citenamefont {Sikes}, \citenamefont {Jiang}, \citenamefont
  {Klima}, \citenamefont {Fudenberg}, \citenamefont {Hone}, \citenamefont
  {Kim},\ and\ \citenamefont {Stormer}}]{bolotin2008}%
  \BibitemOpen
  \bibfield  {author} {\bibinfo {author} {\bibfnamefont {K.}~\bibnamefont
  {Bolotin}}, \bibinfo {author} {\bibfnamefont {K.}~\bibnamefont {Sikes}},
  \bibinfo {author} {\bibfnamefont {Z.}~\bibnamefont {Jiang}}, \bibinfo
  {author} {\bibfnamefont {M.}~\bibnamefont {Klima}}, \bibinfo {author}
  {\bibfnamefont {G.}~\bibnamefont {Fudenberg}}, \bibinfo {author}
  {\bibfnamefont {J.}~\bibnamefont {Hone}}, \bibinfo {author} {\bibfnamefont
  {P.}~\bibnamefont {Kim}}, \ and\ \bibinfo {author} {\bibfnamefont
  {H.}~\bibnamefont {Stormer}},\ }\href@noop {} {\bibfield  {journal} {\bibinfo
   {journal} {Solid State Communications}\ }\textbf {\bibinfo {volume} {146}},\
  \bibinfo {pages} {351} (\bibinfo {year} {2008})}\BibitemShut {NoStop}%
\bibitem [{\citenamefont {Morozov}\ \emph {et~al.}(2008)\citenamefont
  {Morozov}, \citenamefont {Novoselov}, \citenamefont {Katsnelson},
  \citenamefont {Schedin}, \citenamefont {Elias}, \citenamefont {Jaszczak},\
  and\ \citenamefont {Geim}}]{morozov2008}%
  \BibitemOpen
  \bibfield  {author} {\bibinfo {author} {\bibfnamefont {S.~V.}\ \bibnamefont
  {Morozov}}, \bibinfo {author} {\bibfnamefont {K.~S.}\ \bibnamefont
  {Novoselov}}, \bibinfo {author} {\bibfnamefont {M.~I.}\ \bibnamefont
  {Katsnelson}}, \bibinfo {author} {\bibfnamefont {F.}~\bibnamefont {Schedin}},
  \bibinfo {author} {\bibfnamefont {D.~C.}\ \bibnamefont {Elias}}, \bibinfo
  {author} {\bibfnamefont {J.~A.}\ \bibnamefont {Jaszczak}}, \ and\ \bibinfo
  {author} {\bibfnamefont {A.~K.}\ \bibnamefont {Geim}},\ }\href@noop {}
  {\bibfield  {journal} {\bibinfo  {journal} {Phys. Rev. Lett.}\ }\textbf
  {\bibinfo {volume} {100}},\ \bibinfo {pages} {016602} (\bibinfo {year}
  {2008})}\BibitemShut {NoStop}%
\bibitem [{\citenamefont {Stone}\ \emph {et~al.}(2012)\citenamefont {Stone},
  \citenamefont {Downing},\ and\ \citenamefont {Portnoi}}]{stone2012}%
  \BibitemOpen
  \bibfield  {author} {\bibinfo {author} {\bibfnamefont {D.~A.}\ \bibnamefont
  {Stone}}, \bibinfo {author} {\bibfnamefont {C.~A.}\ \bibnamefont {Downing}},
  \ and\ \bibinfo {author} {\bibfnamefont {M.~E.}\ \bibnamefont {Portnoi}},\
  }\href {\doibase 10.1103/PhysRevB.86.075464} {\bibfield  {journal} {\bibinfo
  {journal} {Phys. Rev. B}\ }\textbf {\bibinfo {volume} {86}},\ \bibinfo
  {pages} {075464} (\bibinfo {year} {2012})}\BibitemShut {NoStop}%
\bibitem [{\citenamefont {{Wolfram Research Inc.}}(2010)}]{ram2010}%
  \BibitemOpen
  \bibfield  {author} {\bibinfo {author} {\bibnamefont {{Wolfram Research
  Inc.}}},\ }\href {http://www.wolfram.com} {\emph {\bibinfo {title}
  {Mathematica 8.0}}} (\bibinfo {year} {2010})\BibitemShut {NoStop}%
\end{thebibliography}%

\bibliographystyle{apsrev4-1}

\end{document}